\documentclass[twocolumn,showpacs,preprintnumbers,nofootinbib,prd,superscriptaddress]{revtex4-1}
\usepackage{graphicx,amssymb,amsmath,amsthm,amsfonts,epsfig}

\usepackage[linktocpage]{hyperref}
\usepackage[usenames]{color}
\usepackage{epstopdf}

\def\pa{\partial}

\def\nn{\nonumber}

\newcommand{\ben}{\begin{enumerate}}
\newcommand{\een}{\end{enumerate}}

\def\be{\begin{equation}}
\def\ee{\end{equation}}
\def\bea{\begin{eqnarray}}
\def\eea{\end{eqnarray}}
\newcommand{\beq}{\begin{eqnarray}}
\newcommand{\eeq}{\end{eqnarray}} 

\begin{document}
\title{Slowly Rotating Neutron Stars in Scalar-Tensor Theories}

\author{Paolo Pani}\email{paolo.pani@tecnico.ulisboa.pt}
\affiliation{CENTRA, Departamento de F\'{\i}sica, Instituto Superior T\'ecnico, Universidade T\'ecnica de Lisboa - UTL,
Av.~Rovisco Pais 1, 1049 Lisboa, Portugal.}

\author{Emanuele Berti}\email{eberti@olemiss.edu}
\affiliation{Department of Physics and Astronomy, The University of Mississippi, University, MS 38677, USA.}

\date{\today} 

\begin{abstract} 
We construct models of slowly rotating, perfect-fluid neutron stars by
extending the classical Hartle-Thorne formalism to generic
scalar-tensor theories of gravity. Working at second order in the
dimensionless angular momentum, we compute the mass $M$, radius $R$,
scalar charge ${q}$, moment of inertia $I$ and spin-induced quadrupole
moment $Q$, as well as the tidal and rotational Love numbers.  Our
formalism applies to generic scalar-tensor theories, but we focus in
particular on theories that allow for spontaneous scalarization. It
was recently discovered that the moment of inertia, quadrupole moment
and Love numbers are connected by approximately universal (i.e.,
equation-of-state independent) ``I-Love-Q'' relations.  We find that
similar relations hold also for spontaneously scalarized stars. More
interestingly, the I-Love-Q relations in scalar-tensor theories
coincide with the general relativistic ones within less than a few
percent, even for spontaneously scalarized stars with the largest
couplings allowed by current binary-pulsar constraints. This implies
that astrophysical measurements of these parameters cannot be used to
discriminate between general relativity and scalar-tensor theories,
even if spontaneous scalarization occurs in nature.  Because of the
well known equivalence between $f(R)$ theories and scalar-tensor
theories, the theoretical framework developed in this paper can be
used to construct rotating compact stellar models in $f(R)$
gravity. Our slow-rotation expansion can also be used as a benchmark
for numerical calculations of rapidly spinning neutron stars in
generic scalar-tensor theories.
\end{abstract}

\maketitle

\section{Introduction}
Compact objects such as black holes and neutron stars (NSs) are ideal
astrophysical laboratories to test the strong-field regime of general
relativity (GR)~\cite{Will:2005va,Psaltis:2008bb,Yunes:2013dva}.

The no-hair and uniqueness theorems~\cite{Chrusciel:2012jk} guarantee
that astrophysical black holes in GR are the simplest macroscopic
objects in nature, with structure and dynamics that are determined
only by their mass and spin (but see~\cite{Herdeiro:2014goa} for a recent interesting counterexample). 
Therefore it is relatively easy (at least
conceptually, if not in practice) to detect smoking guns of new
gravitational physics by mapping the multipolar structure of a
black-hole spacetime (see
e.g.~\cite{Ryan:1995wh,Ryan:1997hg,Merritt:2009ex}) or by measuring
the oscillation frequencies of black holes produced as a result of a
compact binary merger~\cite{Berti:2005ys,Berti:2009kk}.

For NSs the situation is qualitatively different because of our poor
understanding of the equation of state (EOS) of high-density nuclear
matter. Different EOSs give rise to very different macroscopic NS
properties, such as masses and radii. The growing wealth of NS
observations holds great promise to constrain the
EOS (cf.~\cite{Lattimer:2006xb,Ozel:2010fw,Steiner:2010fz,Psaltis:2013fha} and \cite{2013arXiv1312.0029M} for a recent review),
but the degeneracy between different EOS models and strong-field
gravitational physics limits our ability to carry out tests of
strong-field gravity. The reason is that uncertainties in our
knowledge of the EOS are typically much larger than putative
corrections from extensions of GR that are theoretically viable and
pass weak-field tests.

This state of affairs has changed after the discovery by Yagi and
Yunes~(\cite{Yagi:2013bca,Yagi:2013awa}; see also
\cite{Stein:2013ofa}) that suitable dimensionless combinations of the
moment of inertia $I$, the tidal Love number $\lambda$ and the
spin-induced quadrupole moment $Q$ of slowly-rotating NSs satisfy
largely universal relations, where by ``universal'' we mean that these
relations do not depend on the NS EOS within an accuracy of a few
percent~\cite{Lattimer:2012xj}. The universality is remarkably robust:
various investigations showed that universal relations apply in GR
also when the star rotates
rapidly~\cite{Doneva:2013rha,Pappas:2013naa,Chakrabarti:2013tca,Yagi:2014bxa},
for moderately strong magnetic fields~\cite{Haskell:2013vha}, and for
stars whose parameters evolve dynamically due to interactions with a
companion~\cite{Maselli:2013mva}. Various other nearly universal
relations involving NSs have been discussed in the literature, and our
understanding of the nature of these relations is steadily
improving~\cite{Lattimer:2000nx,Tsui:2004qd,Urbanec:2013fs,Baubock:2013gna,Yagi:2013sva,AlGendy:2014eua}.
The I-Love-Q relations are interesting for astrophysics because, if we
assume that GR provides an accurate description of the
strong-curvature regime, current and future observational facilities
(e.g. ATHENA+~\cite{Barret:2013bna}, LOFT~\cite{Bozzo:2013txa}, NICER
\cite{2012SPIE.8443E..13G} and the SKA~\cite{Smits:2008cf} in the
electromagnetic spectrum, as well as Advanced LIGO~\cite{AdvLIGO},
Advanced Virgo~\cite{AdvVirgo}, KAGRA~\cite{KAGRA} and the Einstein
Telescope~\cite{Punturo:2010zz} in the gravitational-wave spectrum)
may allow us to infer all three I-Love-Q quantities from the
measurement of a single element of the triad (either $I$, $Q$ or
$\lambda$).

The existence of EOS-independent relations between the macroscopic
parameters of compact stars in GR allows us, at least in principle, to
circumvent the EOS-degeneracy problem mentioned above in the context
of tests of strong-field gravity.  Yagi and Yunes proposed the
interesting possibility to constrain the underlying theory of gravity
from measurements of the ``no-hair like'' I-Love-Q
relations~\cite{Yagi:2013bca,Yagi:2013awa}: if these relations are
{\em different} in alternative theories of gravity (yet
EOS-independent within each theory), then precision measurements of
two of these quantities may allow us to discriminate between GR and
possible extensions of the theory.

This is one of the most interesting applications of the I-Love-Q
relations, but so far it has been explored only for two proposed
alternatives to GR: Dynamical Chern-Simons (DCS)
gravity~\cite{Alexander:2009tp} and Eddington-inspired Born-Infeld
(EiBI) gravity~\cite{Banados:2010ix,Pani:2011mg,Pani:2012qb}.  For DCS
gravity, it has been shown that tests based on the I-Love-Q relations
can potentially constrain the theory better than current experimental
bounds, basically because binary pulsar bounds on the theory are not
very stringent~\cite{Yagi:2013awa}. On the other hand, the I-Love-Q
relations in EiBI gravity were shown to be degenerate with their GR
counterparts~\cite{Sham:2013cya}. This degeneracy is interesting, but
not surprising.  EiBI gravity does not contain any extra degree of
freedom with respect to GR. Solutions of the stellar structure
equations in GR can be mapped to solutions in EiBI theory with an
effective EOS~\cite{Delsate:2012ky} that is only slightly different
from the corresponding GR EOS, given current experimental constraints
on EiBI theory. For this reason, the indistinguishability of GR and
EiBI theory is conceptually almost trivial. Furthermore there are
issues with EiBI gravity, because the theory shares several of the
pathologies that affect Palatini $f(R)$
theories~\cite{Barausse:2007pn}, including curvature singularities at
the surface of polytropic stars and a problematic Newtonian
limit~\cite{Pani:2012qd}.

In this work we investigate one of the most natural (and certainly the
best studied) extensions of GR, namely scalar-tensor gravity
\cite{Fujii:2003pa,Will:2005va}. This is a fundamental theory with a
well defined initial value problem~\cite{Salgado:2007ep} where gravity
is mediated by the usual massless graviton and by a fundamental scalar
field.  The historical development of scalar-tensor theories was
driven by a desire to investigate the role of Mach's principle in
gravity, but scalar degrees of freedom are ubiquitous in high-energy
extensions of Einstein's theory~\cite{Sotiriou:2014jla}, in models
that try to explain cosmological observations via modified
gravity~\cite{Clifton:2011jh} and in inflation scenarios
\cite{Martin:2013tda}. Certain classes of scalar-tensor theories are
equivalent to $f(R)$
gravity~\cite{Sotiriou:2008rp,DeFelice:2010aj}. Furthermore,
scalar-tensor gravity can be considered as a simple phenomenological
proxy for more complex strong-field extensions of GR.

In the context of NS physics, the interest in scalar-tensor gravity
was revived after certain scalar-tensor theories were shown to produce
``spontaneous scalarization''~\cite{Damour:1993hw,Damour:1996ke}. In a
nutshell, these theories allow for the same NS solutions as in GR, but
the GR solutions become unstable beyond a critical central pressure and
--~in a phase transition akin to ferromagnetism~-- other solutions
with a nonzero scalar charge appear. These ``spontaneously
scalarized'' solutions are stable and can display relatively large
deviations from their GR counterparts, even if the theory passes all
weak-field tests~\cite{Damour:1996ke}. This interesting phenomenon has
been recently shown to be strengthened in dynamical situations, such
as the final stages of a binary NS
merger~\cite{Barausse:2012da,Shibata:2013pra,Palenzuela:2013hsa}, and
it has been shown to occur also for black holes surrounded by
matter~\cite{Cardoso:2013fwa,Cardoso:2013opa}. 

Doneva {\em et al.} \cite{Doneva:2013qva} recently studied scalarized
configurations for rapidly rotating stars, showing that rotation
enhances the effects of scalarization. The present paper is
complementary to their work: we adopt the slow-rotation approximation
(rather than solving the Einstein equations numerically for arbitrary
rotation), but we extend the work of \cite{Doneva:2013qva} by
extracting all relevant physical quantities, including the quadrupole
moment and the Love numbers, at second order in the slow-rotation
expansion. 

Our main result is that experimentally viable, spontaneously
scalarized NS solutions have the {\em same} I-Love-Q relations as GR
solutions within a few percent, i.e. the modified-gravity corrections
are degenerate with the (small) deviations from universality within
GR.  Therefore, experimental measurements of the I-Love-Q relations
cannot be used to distinguish GR from scalar-tensor theories, nor to
put constraints on the latter that are more stringent than those
currently in place~\cite{Freire:2012mg}. 

These results, together with those for the very special case of EiBI
gravity~\cite{Sham:2013cya}, suggest that --~for most theories that
are well constrained by weak-field tests~-- the modified I-Love-Q
relations might be indistinguishable from their GR counterpart.
On the other hand, our study proves that the I-Love-Q universality is
remarkably robust even against beyond-GR corrections: as long as the
modifications to GR affect both the strong- and weak-field regimes
(and can therefore be strongly constrained by weak-field experiments),
a measurement of one element of the triad can be used to infer the
remaining two quantities within a few percent.

The paper is organized as follows. In Sec.~\ref{sec:framework} we
present the main ingredients of our formalism to construct
slowly-rotating NS configurations to second order in rotation in
generic scalar-tensor theories of gravity. In Sec.~\ref{sec:results}
we focus on a theory that allows for spontaneous scalarization and we
present our numerical results, showing that the universal I-Love-Q
relations are very close to their GR counterparts for theories that
are compatible with binary pulsar experiments. In
Sec.~\ref{sec:conclusions} we summarize the implications and possible
extensions of our work. In Appendix \ref{app:eqs} we present the field
equations in the Einstein frame, and in Appendix~\ref{app:EtoJ} we
discuss how to relate physical quantities in the Jordan frame to
quantities computed in the Einstein frame.

\section{Framework}\label{sec:framework}
A generic class of scalar-tensor theories in the Jordan frame is
described by the action~\cite{Fujii:2003pa}
\bea
\label{actionST}
\tilde{S}&=&\int d^4x \frac{\sqrt{-\tilde g}}{16\pi G}\left(F(\tilde \phi)\tilde R-Z(\tilde \phi)\tilde g^{\mu\nu}\partial_{\mu}\tilde \phi\partial_{\nu}\tilde \phi-U(\tilde \phi)\right)\nonumber\\
&+&S_m(\Psi_m;\tilde g_{\mu\nu})\,,
\eea
where $\tilde R$ is the Ricci scalar constructed out of the spacetime
metric $\tilde g_{\mu\nu}$, $\tilde \phi$ is a scalar field, and
$\Psi_m$ collectively denotes the matter fields (which are minimally
coupled to $\tilde g_{\mu\nu}$). The constant $G$ is related to the
physical gravitational constant (as measured in a Cavendish-type
experiment), and from now on we will set it to unity together with the
speed of light (see Appendix~\ref{app:EtoJ} for more details). Here
and below we denote by a tilde quantities defined in the Jordan
frame. Choosing the functions $F$, $Z$ and $U$ determines a specific
theory within the class, up to a degeneracy due to the freedom to
redefine the scalar~\cite{Sotiriou:2007zu}.

By performing the transformations
\begin{eqnarray}
 {g}_{\mu\nu}&=&F(\tilde \phi)\tilde g_{\mu\nu}\,,\quad A({\Phi})=F^{-1/2}({\tilde\phi})\,,\quad V({\Phi})=\frac{U({\tilde\phi})}{F^2({\tilde\phi})}\,,\nn\\
 {\Phi}({\tilde\phi})&=&\int  \frac{d{\tilde\phi}}{\sqrt{4\pi}}\,\sqrt{\frac{3}{4}\frac{F'({\tilde\phi})^2}{F({\tilde\phi})^2}+\frac{1}{2}\frac{Z({\tilde\phi})}{F({\tilde\phi})}}\,,
\end{eqnarray}
the theory can be recast in the so-called Einstein frame, where the
action reads
\bea
S&=&\int d^4x \sqrt{-{ g}}\left(\frac{{R}}{16\pi}-\frac{1}{2}{ g}_{\mu\nu}\partial^\mu{\Phi}\partial^\nu{\Phi}-\frac{V({\Phi})}{16\pi}\right)\nonumber\\
&+&S(\Psi_m;A({\Phi})^2 g_{\mu\nu})\,.\label{actionEinstein}
\eea
In the Einstein frame the scalar field is minimally coupled to
gravity, but the matter fields $\Psi_m$ are minimally coupled to the
metric $\tilde g_{\mu\nu}\equiv A({\Phi})^2 g_{\mu\nu}$, and
nonminimally coupled to the conformal Einstein metric $g_{\mu\nu}$.
The field equations in the Einstein frame read
\begin{eqnarray}
 G_{\mu\nu}&=&8\pi T_{\mu\nu}\nn\\
 &+&8\pi\left(\partial_\mu{\Phi}\partial_\nu{\Phi}-\frac{ g_{\mu\nu}
}{2}\partial_\sigma{\Phi}\partial^\sigma\Phi\right)-\frac{ g_{\mu\nu}}{2} 
V({\Phi})\,,\label{einein}\\
 \square{\Phi}&=&-\frac{A'({\Phi})}{A({\Phi})} T+\frac{V'({\Phi})}{16\pi}\,,\label{einscalar}
\end{eqnarray}
where the Einstein-frame stress-energy tensor is related to the
physical (Jordan-frame) stress-energy tensor by
\begin{equation}
 {T^\mu_\nu}=A^4({\Phi}) \tilde T^{\mu}_{\nu}\,,\quad  T_{\mu\nu}=A^2({\Phi}) \tilde T_{\mu\nu}\,,\quad T=A^4({\Phi}) \tilde T\,,\nn
\end{equation}
and the Jordan-frame stress-energy tensor for a perfect-fluid
reads
\begin{equation}
 \tilde T^{\mu\nu}=\left(\rho+P\right)\tilde u^\mu\,\tilde u^\nu+\tilde g^{\mu\nu}P\,.	\label{Tmunu_fluid}
\end{equation}
We omit a tilde on the Jordan-frame pressure $P$, density $\rho$ and
fluid angular velocity $\Omega$, but since we only consider these
quantities in the Jordan frame, the notation should not be
ambiguous. To
second order in $\Omega$, the fluid four-velocity reads $\tilde
u^\mu=(\tilde u^0,0,0,\epsilon\Omega \tilde u^0)$, where
\begin{equation}
 \tilde u^0=\left[-(\tilde g_{tt}+2\epsilon\Omega \tilde g_{t\varphi}+\epsilon^2\Omega^2 \tilde g_{\varphi\varphi})\right]^{-1/2}\,
\end{equation}
and $\epsilon$ is a bookkeeping slow-rotation parameter. In this paper
all physical quantities characterizing the structure of a compact star
will be expanded to ${\cal O}(\epsilon^2)$.  Note that
\begin{equation}
 {T}_{\mu\nu}=A^4(\Phi){g}_{\mu \sigma} g_{\nu \tau}\left[\left(\rho+P\right)u^\sigma\,u^\tau+g^{\sigma\tau}P\right]\,,\label{Tmunu_fluid_Einstein}
\end{equation}
where, in the Einstein frame, $u^\mu=(u^0,0,0,\epsilon\Omega u^0)$ with
\begin{equation}
  u^0=\left[-( g_{tt}+2\epsilon\Omega g_{t\varphi}+\epsilon^2\Omega^2 g_{\varphi\varphi})\right]^{-1/2}\,,
\end{equation}
so that $u^\mu=A(\Phi) \tilde u^\mu$. Because of the transformation of $u^\mu$, the fluid angular velocity $\Omega$ is the
same in both frames: $\Omega\equiv
u^\varphi/u^t=(A(\Phi)\tilde{u}^\varphi)/(A(\Phi)\tilde{u}^t)$.

Following Hartle and Thorne~\cite{Hartle:1967he,Hartle:1968si}, the
most general stationary axisymmetric metric $g_{\mu\nu}$ to ${\cal
  O}(\epsilon^2)$ in rotation can be written as
\begin{eqnarray}
 ds^2&&=-e^\nu\left[1+2\epsilon^2\left(h_0+h_2 P_2\right)\right]dt^2\nn\\
 &&+\frac{1+2\epsilon^2(m_0+m_2P_2)/(r-2m)}{1-2m/r}dr^2\nn\\
 &&+r^2\left[1+2\epsilon^2(v_2-h_2)P_2\right]\left[d\vartheta^2+\sin^2\vartheta(d\varphi-\epsilon\omega dt)^2\right]\,,\nn\\ \label{metric}
\end{eqnarray}
where $P_2=P_2(\cos\vartheta)=(3\cos^2\vartheta-1)/2$ is a Legendre
polynomial. The radial functions $\nu$ and $m$ are of zeroth order in
rotation, $\omega$ and the related quantity $\bar\omega=\Omega-\omega$
(that will be useful below) are of first order, and $h_0$, $h_2$,
$m_0$, $m_2$, $v_2$ are of second order.
Under an infinitesimal rotation the scalar field, the pressure and the
density all transform as scalars. As shown in~\cite{Hartle:1967he,Hartle:1968si}, in order to perform a valid perturbative expansion it is necessary to transform the radial coordinate in such a way that the deformed density in the new coordinates coincides with the unperturbed density at the same location. It can be shown that this transformation is formally equivalent to working in the original coordinates but expanding the pressure and the density as
\begin{eqnarray}
 P&=&P_0+\epsilon^2(\rho_0+P_0)(p_0+p_2 P_2)\,,\\
 \rho&=&\rho_0+\epsilon^2(\rho_0+P_0)\frac{\pa\rho_0}{\pa P_0}(p_0+p_2 P_2)\,,\label{rho}
\end{eqnarray}
where we have assumed a barotropic EOS of the form
$P=P(\rho)$. On the other hand, the scalar field is not affected by the fluid displacement and is simply expanded as
\begin{equation}
  {\Phi}=\Phi_0+\epsilon^2(\phi_0+\phi_2 P_2)\,. \label{scalardef}
\end{equation}
By plugging this decomposition into the
gravitational and scalar-field
equations~\eqref{einein}--\eqref{einscalar} and by solving the
equations order by order in $\epsilon$ we obtain a system of ordinary
differential equations (ODEs).

We could in principle include a nonzero potential $V(\Phi)$ in these
equations. While the inclusion of the potential is crucial in a
cosmological context and can affect binary
dynamics~\cite{Alsing:2011er}, for the present study of isolated
compact objects we will assume that the scalar-field mass (and other
self-interactions described by the potential) are small enough to be
negligible, and we will focus on the case $V(\Phi)\equiv0$. The final
form of the equations when $V(\Phi)\equiv0$ is given in
Appendix~\ref{app:eqs}. The more general equations for $V(\Phi)\neq
0$, along with the procedure to integrate the equations numerically
and extract the relevant quantities (discussed below), are presented
in a publicly available {\scshape Mathematica}
notebook~\cite{webpages}.

\subsection{Integration of the field equations and extraction of the moment of inertia and quadrupole moment}

The perturbation equations are very lengthy, and we summarize them in
Appendix~\ref{app:eqs}. Schematically, the system can be written in
the form
\begin{equation}
 \frac{d\boldsymbol{Y}(r)}{dr}=\boldsymbol{A}\boldsymbol{Y}(r)\,, \label{system}
\end{equation}
where 
\begin{eqnarray}
 \boldsymbol{Y}(r)&=&\left\{m,P_0,\nu,\Phi_0,\Phi_0',\bar\omega,\bar\omega',m_0,p_0,h_0,v_2,h_2,\right.\nn\\
 &&\left.v_2^{(h)},h_2^{(h)},\phi_0,\phi_0',\phi_2,\phi_2',\phi_0^{(h)},{\phi_0^{(h)}}',\phi_2^{(h)},{\phi_2^{(h)}}'\right\}\,,\nn\\
\label{eq:variables}
\end{eqnarray}
and $\boldsymbol{A}$ is a 22-dimensional square matrix. The functions $m_2$ and $p_2$ are algebraically related to the others.
This system of linear equations must be solved by imposing regularity
at the center of the star, continuity at the surface, and asymptotic
flatness at infinity. As we will discuss shortly, when we work at
second order in rotation some of the equations are inhomogeneous, and
the boundary conditions at infinity can be conveniently imposed using
the corresponding homogeneous solutions, denoted by a superscript
``(h)'' in Eq.~(\ref{eq:variables}). This is the reason why we
integrate the full system~\eqref{system}, including both inhomogeneous
and homogeneous quantities. The problem can be solved order by order
in $\epsilon$, but in practice it is more convenient to integrate all
(zeroth-, first- and second-order) equations simultaneously.

Near the center, the leading-order behavior of the regular solution is
\begin{center}
 \begin{tabular}{lll}
 $m \sim m_3 r^3\,,\quad$	& $\nu\sim \nu_c\,,\quad$	& $P_0\sim P_{c}\,,\quad$ \\
 $\Phi_0\sim \Phi_{c}\,,\quad$	& $\bar\omega\sim\bar\omega_c\,,\quad$ &\\
 $p_0\sim p_{02}r^2\,,\quad$	& $v_2\sim v_{22}r^2\,,$   & $h_2\sim h_{24}r^4\,,$\\
 $v_2^{(h)}\sim v_{22}^{(h)}r^2\,,$   & $h_2^{(h)}\sim h_{24}^{(h)}r^4\,,\quad$ & $\phi_0\sim\phi_{04}r^4$\,,\\
 $\phi_2\sim\phi_{22}r^2\,,\quad$   &  $\phi_0^{(h)}\sim\phi_{00}^{(h)}\,,\quad$   &  $\phi_2^{(h)}\sim\phi_{22}^{(h)}r^2$\,,
\end{tabular}
\end{center}
where $P_{c}$, $\nu_c$, $\Phi_c$ and $\bar\omega_c$ denote the values of
the corresponding functions at the center of the star.
Not all of the series expansion coefficients listed above are
independent. Furthermore, to improve numerical stability, in our
{\scshape Mathematica} notebook we included higher-order terms in the
series expansions near the center. Without loss of generality, $\nu_c$
can be set to unity through a time rescaling, and $\bar\omega_c$ can
be set to unity by using the fact that the relevant ODE,
Eq.~\eqref{eqomega}, is homogeneous in $\bar\omega$. Since the
gyromagnetic factor is linear in $\Omega$, at first order in rotation
the entire family of spinning solutions can be obtained from a single
element of the family by a suitable rescaling. Therefore, the
first-order equilibrium structure is defined by two parameters: the
central pressure $P_{c}$ and the central value of the scalar field
$\Phi_{c}$.

Using the boundary conditions above, the system~\eqref{system} can be
integrated from $r=0$ to the stellar surface $r=R$, defined by
$P_0(R)=0$. By imposing continuity at the surface, all dynamical
variables can be computed at $r=R$, and the system can be integrated
outwards from $r=R$ to infinity in the vacuum exterior, where
$P_0(r)=\rho_0(r)=0$. To zeroth order the quantity $\Phi_{c}$ is fixed
by requiring asymptotic flatness, i.e.
\begin{eqnarray}\label{Cdef}
 m(r)&\to&M\,,\quad  \nu(r)\to 0\,,\quad \Phi_0(r)\to \Phi_0^\infty+\frac{C}{r}\,,
\end{eqnarray}
as $r\to\infty$, where $M$ is the mass of the
star\footnote{Since we integrate the equations derived from the
  action~\eqref{actionEinstein}, all quantities are here defined in
  the Einstein frame. When presenting the results, however, we shall
  consider the corresponding quantities in the physical (Jordan)
  frame.  The relevant transformations are discussed in
  Appendix~\ref{app:EtoJ}. Strictly speaking, only quantities in the
  Jordan frame can be observationally interpreted as the total mass,
  charge, angular momentum, etcetera: for example, the Jordan-frame
  mass $\tilde{M}$ is the quantity measured by applying
  Kepler's law to weak-field orbits.} and $C={q}/\sqrt{4\pi}$,
where ${q}$ is the scalar charge of the star. For any fixed value
of $P_{c}$, we can determine the central scalar field $\Phi_{c}$
that enforces asymptotic flatness through a shooting procedure.
The angular momentum $J$ of the star is related to the asymptotic
expansion of $\bar\omega$ as follows:
\begin{equation}
 \bar\omega\to \Omega-\frac{2 J}{r^3}+\frac{12\pi J C^2}{5 r^5}\quad {\rm as} \quad r\to\infty\,.
\end{equation}
The moment of inertia of the star $I$ is simply
\begin{equation}
 I=\frac{J}{\Omega}\,,
\end{equation}
and at leading order in the slow-rotation expansion it is independent
of the NS spin (see~\cite{Benhar:2005gi,Yagi:2014bxa} for higher-order
corrections).

Let us now discuss the boundary conditions for quantities of second
order in rotation. Generic solutions of the inhomogeneous system are
irregular at infinity, the general behavior being
\begin{equation}
 h_2(r)\to h_2^{\rm irr}r^2\,,\qquad \phi_2(r)\to \phi_2^{\rm irr}r^2\,, \label{divergent}
\end{equation}
where $h_2^{\rm irr}$ and $\phi_2^{\rm irr}$ are constants.
The regular solution can be constructed by a suitable linear
combination of a particular inhomogeneous solution and the
corresponding homogeneous solution. The homogeneous system forms a
two-parameter family, defined by the values of
$(h_2^{(h)},\,\phi_2^{(h)})$ at the center. Without loss of
generality, we construct two linearly independent solutions of the
homogeneous system by choosing the values $(1,\,0)$ and $(0,\,1)$ for
these parameters. A linear combination of these solutions is added to
a particular solution of the inhomogeneous problem, and we choose the
coefficients of the linear combination in order to cancel the
divergent terms~\eqref{divergent}. A similar procedure is employed in
the GR case~\cite{Hartle:1968si}. The leading-order, large-distance
behavior of the regular solutions reads
\begin{equation}
 h_2(r)\to \frac{Q}{r^3}\,,\qquad \phi_2(r)\to \frac{Q_s}{r^3}\,,
\end{equation}
where $Q$ is the spin-induced quadrupole moment of the
star~\cite{Hartle:1968si} and $Q_s$ is a new quantity related to a
quadrupolar deformation of the scalar field.
A similar procedure is applied to compute the regular solutions $m_0$
and $\phi_0$, whose asymptotic behavior reads $m_0\to\delta M$ and
$\phi_0\to \delta{q}/(\sqrt{4\pi}r)$, where $\delta M$ and $\delta{q}$
are the second-order corrections to the total mass and to the total
scalar charge, respectively.

\subsection{Tidal Love numbers in scalar-tensor gravity}
We compute the tidal Love numbers in scalar-tensor gravity by
extending the relativistic formalism developed by Hinderer in
GR~\cite{Hinderer:2007mb}, which in turn is based on the analysis of
metric perturbations sourced by an external quadrupolar tidal
field~\cite{Thorne:1967}.

Restricting the analysis to $l=2$, static, even-parity perturbations,
a consistent ansatz is obtained from a subset of the decomposition in
Eqs.~\eqref{metric}--\eqref{rho} by setting
$\bar\omega_1=h_0=m_0=p_0=\phi_0=0$. After redefining $h_2=-H_0/2$,
$m_2=(1-2m/r)r H_2/2$ and $\phi_2=\Phi_2$, it can be shown that the
field equations imply $H_2=H_0$, and that the perturbation equations
reduce to a coupled system of second-order ODEs:
\begin{eqnarray}
 H_0''+c_1 H_0'+c_0 H_0&=&c_s \phi_2\,,    \label{eqlove1}\\
 \Phi_2''+d_1 \Phi_2'+d_0 \Phi_2&=&d_s H_0 \label{eqlove2}\,,
\end{eqnarray}
with 
\begin{eqnarray}
 c_1&&=d_1=\frac{1+e^{\Lambda } \left[1+4 \pi  r^2 \left(A^4 (P_0-\rho_0)-2V\right)\right]}{r}\,,\\
 c_0&&=-\frac{1}{ r^2}\left[ e^{2 \Lambda } \left(1+8 \pi  r^2 \left(A^4 P_0-V\right)\right)^2+ \left(1-4 \pi  r^2 \Phi_0'^2\right)^2\right.\nn\\
 &&\left.+4 e^{\Lambda } \left(1+\pi  r^2 \left(8  V+2  \left(1-8 \pi  r^2 V\right) \Phi_0'^2\right.\right.\right.\nn\\
 &&\left.\left.\left.-A^4 \left(\frac{\rho_0}{P_\rho} +5 \rho_0 +P_0 \left(\frac{1}{P_\rho}+13 -16 \pi  r^2 \Phi_0'^2\right)\right)\right)\right)\right]\,,\nn\\ \\
 d_0&&=e^{\Lambda }  \left[A^2 A'^2 \left(\frac{P_0+\rho_0}{P_\rho}+6(P_0-\rho_0) \right)+A^3 (3 P_0-\rho_0 ) A''\right. \nn\\
 &&\left.- V''\right]-\frac{6}{r^2}e^{\Lambda }-16  \pi \Phi_0'^2\,,\\
 c_s&&= 16\pi d_s=\frac{2\Phi_0'}{r}  \left(4 \pi  r^2 \Phi_0'^2-1\right) \nn\\
 &&-8\pi e^{\Lambda }\left(\frac{A^3 A'}{P_\rho} ((9 P_\rho-1) P_0+(P_\rho-1) \rho_0 ) -2 V'\right.\nn\\
 &&\left.+\frac{2}{r}  \left(1+8 \pi  r^2 \left(A^4 P_0-V\right)\right) \Phi_0'\right)\,,
\end{eqnarray}
where $P_\rho\equiv \partial P_0/\partial \rho_0$,
$e^{-\Lambda(r)}=1-2m(r)/r$ and primes denotes derivatives with
respect to the argument, i.e. $H_0'\equiv dH_0/dr$ and $A'\equiv
dA/d\Phi_0$. As discussed in Appendix~\ref{app:eqs}, the variables $p_2$ and $v_2$ can be determined in terms of the other functions.
As expected, the nonhomogeneous coefficients of the
differential system depend only on zeroth-order (in $\Omega$)
background quantities, so that Eqs.~\eqref{eqlove1} and
\eqref{eqlove2} can be solved together with the original
system~\eqref{system}, considering $H_0$ and $\Phi_2$ as additional
independent variables.

The regular solutions of Eqs.~\eqref{eqlove1}--\eqref{eqlove2} near
the center of the star behave as $H_0\sim H_{02}r^2$, $\Phi_2\sim
\Phi_{22}r^2$, and higher order coefficients can be expressed in terms
of $H_{02}$ and $\Phi_{22}$ by solving the equations order by order
near the center. We integrate the system twice with boundary
conditions $(H_{02},\Phi_{22})=(1,0)$ and $(H_{02},\Phi_{22})=(0,1)$
at the center, respectively. By imposing continuity of $H_2$, $\Phi_2$
and their derivatives at the radius we construct two linearly
independent solutions in the entire domain. Finally, we construct a
linear combination of these solutions such that $\Phi_2$ (as obtained
by the linear combination) is regular at infinity, i.e. we impose the
following asymptotic behavior for the linear combination of the two
solutions:
\begin{eqnarray}
    H_0 &\to& a_{-2}r^2+a_{-1}r+a_0 +\frac{a_1}{r}+\frac{a_2}{r^2}+\frac{a_3}{r^3}\,, \label{H0inf}\\
 \Phi_2 &\to& b_0+\frac{b_3}{r^3} \label{Phi2inf}\,,
\end{eqnarray}
where the $a_i$'s and $b_i$'s are constants which can be expressed in
terms of four independent parameters by using the field equations. The
expression above is valid for $V(\Phi)\equiv0$, but it can be easily
generalized to include a scalar potential.  Finally, the tidal Love
number is defined as~\cite{Hinderer:2007mb}
\begin{equation}
 \lambda=\frac{a_3}{3 a_{-2}}\,.
\end{equation}
We note that $a_{-1}=-2Ma_{-2}$ and that $a_0$, $a_1$, $a_2$ and $b_0$
are vanishing when the scalar charge ${q}=0$. Therefore it is harder
to extract the subdominant coefficient $a_3$ from a numerical solution
when ${q}\neq0$. Furthermore, the formalism allows us to extract also
$b_0\propto{q}$ and $b_3$, which is related to a quadrupolar
deformation of the background scalar field. In analogy with the tidal
Love number introduced above, we can define a \emph{scalar} Love
number (with dimensions of $M^3$) as follows:
\begin{equation}
 \lambda_s=\frac{b_3}{b_{0}}\,.
\end{equation}
%

\subsection{$\bar{I}-\bar{\lambda}-\bar{Q}$ relations and the slow-rotation approximation}
Yagi and Yunes~\cite{Yagi:2013bca,Yagi:2013awa} discovered that,
within GR, the dimensionless quantities
\begin{equation}\label{ILoveQbar}
 \bar{I}=\frac{I}{M^3}\,,\qquad  \bar{\lambda}=\frac{\lambda}{M^5}\,,\qquad \bar{Q}=\frac{Q}{M^3\chi^2}\,,
\end{equation}
($\chi=J/M^2$ being the dimensionless spin) satisfy nearly
universal relations that are insensitive to the NS EOS
within an accuracy of the order of a few percent.

Another relevant quantity in the context of universal relations is the
rotational Love number $\lambda^{\rm rot}$, which is related to the
deformability of the NS away from sphericity due to its own
rotation~\cite{Mora:2003wt,Berti:2007cd}. The dimensionless quantity
associated with this number can be expressed in terms of $\bar{I}$ and
$\bar{Q}$ as~\cite{Yagi:2013awa}
\begin{equation}
 \bar{\lambda}^{\rm rot}\equiv\frac{\lambda^{\rm rot}}{M^5}=\bar{I}^2\bar{Q}\,.\label{lambdarot}
\end{equation}

Our stellar structure equations correctly reduce to their GR
counterparts~\cite{Hartle:1968si,Hinderer:2009ca} when $A(\Phi)\equiv 1$ and
$V(\Phi)\equiv0$, and we have tested our code by reproducing the
results of Refs.~\cite{Yagi:2013bca,Yagi:2013awa} in the GR case.
As an additional test we have reproduced the results of
Ref.~\cite{Berti:2004ny} for the mass, radius, moment of inertia and
quadrupole moment for several EOS models within GR.

It is important to remark that the universality discovered
in~\cite{Yagi:2013bca,Yagi:2013awa} can be affected both by
observational uncertainties and by the slow-rotation
approximation. 

The normalizations in Eq.~\eqref{ILoveQbar} involve powers of the mass
$M$ of a {\em nonrotating} star (for $I$ and $\lambda$) as well as
powers of the dimensionless spin $\chi$ (for $Q$). Astrophysical
observations yield unbarred quantities, which must be
normalized by the {\em measured} mass and (dimensionless) spin in
order to satisfy EOS-independent relations. In the second-order
slow-rotation approximation used here, the observable mass $M_{\rm
  rot}$ of a rotating NS is related to the mass of the nonrotating
model by
\be\label{deltaM}
M_{\rm rot} = M + \epsilon^2 \delta M\,.
\ee
where $\epsilon$ is the slow-rotation expansion parameter introduced
earlier. The applicability of the universality relations to infer
(say) $I$ and $Q$ from $\lambda$ will be limited in practice by
measurement errors on $M_{\rm rot}$ and $\lambda$, not by the
remarkably small dispersion between ``barred'' quantities. Since the
normalization involves high powers of $M$, small (say $\sim 5\%$)
errors on the mass would translate into large ($\sim 25\%$) errors on
$\bar \lambda$. Similar considerations apply to the quadrupole moment,
where the normalization is affected by both mass {\em and} spin
measurement errors.

A related limitation in the practical use of I-Love-Q relations is
that universality is effectively broken by a rotation-dependent term,
because $M_{\rm rot}\neq M$. For typical nuclear-physics motivated EOS
and in the mass range of interest for NSs, $\delta M/M \sim 0.3$
(cf. e.g. \cite{Berti:2004ny}, or Fig.~\ref{fig:structure} below). For
millisecond pulsars the small rotation parameter can be as large as
$\epsilon\sim 0.5$ \cite{Hessels:2006ze}, so the difference between
the rotating and nonrotating mass would introduce corrections to the
universality which are of order $(0.5)^2\times 0.3\sim 7.5\%$, larger
than the dispersion in the I-Love-Q relations themselves. While
important in principle, this limitation is not of much concern in
practice, for two reasons: 

\begin{itemize}
\item[1)] The systems for which I-Love-Q tests would be
  astrophysically interesting include double pulsars, where
  precessional effects could lead to measurements of the moment of
  inertia $I$ \cite{Lattimer:2004nj}, and compact binaries
  coalescences that may be observed by future gravitational-wave
  interferometers, allowing for measurements of the tidal Love number
  \cite{Hinderer:2009ca}. As argued convincingly in
  \cite{Yagi:2013bca,Yagi:2013awa}, these systems typically involve
  NSs for which rotation rates are rather low.

\item[2)] More recent studies~\cite{Doneva:2013rha,Pappas:2013naa,Chakrabarti:2013tca,Yagi:2014bxa} show that, at least in GR, the I-Love-Q
  universality is remarkably robust even for fast rotating stellar
  models, when the various quantities are normalized by powers of the
  appropriate (measurable) mass.
\end{itemize}

\begin{figure*}[t]
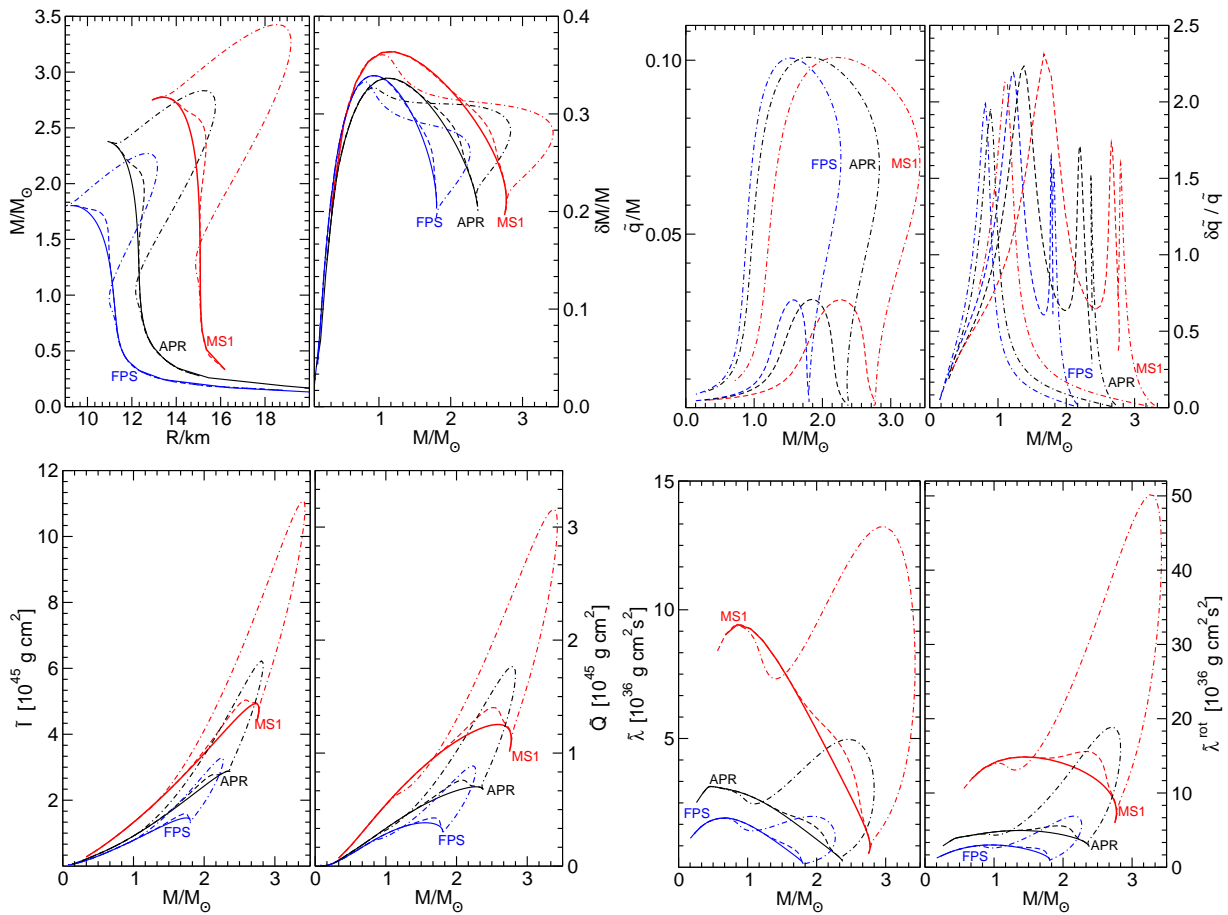

\begin{center}
\begin{tabular}{ll}
\epsfig{file=MR_Jordan.eps,width=8cm,angle=0,clip=true}&
\epsfig{file=CM_Jordan.eps,width=8cm,angle=0,clip=true}\\
\epsfig{file=I-Q-M_Jordan.eps,width=8cm,angle=0,clip=true}&
\epsfig{file=lambda-lambdaR-M_Jordan.eps,width=8cm,angle=0,clip=true}\\
\end{tabular}
\caption{NS configurations in GR (solid lines) and in two
  scalar-tensor theories defined by Eq.~\eqref{actionEinstein} with
  $A(\Phi)\equiv e^{\frac{\beta}{2}\Phi^2}$ and
  $V(\Phi)\equiv0$. Dashed lines refer to $\beta/(4\pi)=-4.5$,
  $\Phi_0^\infty=10^{-3}$; dash-dotted lines refer to
  $\beta/(4\pi)=-6$, $\Phi_0^\infty=10^{-3}$. Each panel shows results
  for three different EOS models (\texttt{FPS}, \texttt{APR}
  and \texttt{MS1}).
  Top-left panel, left inset: relation between the nonrotating mass
  $M$ and the radius $R$. In all plots $M$ refers to the
  Arnowitt-Deser-Misner mass in the Einstein frame: see Appendix
  \ref{app:EtoJ} for a discussion. Top-left panel, right inset:
  relative mass correction $\delta M/M$ induced by rotation as a
  function of the mass $M$ of a nonspinning star with the same central
  energy density [cf.~Eq.~\eqref{deltaM}].
  Top-right panel, left inset: scalar charge $\tilde{q}/M$ as a function
  of $M$. Top-right panel, right inset: relative correction to the
  scalar charge $\delta \tilde{q}/\tilde{q}$ induced by rotation as a
  function of $M$.
  Bottom-left panel: Jordan-frame moment of inertia $\tilde{I}$ (left
  inset) and Jordan-frame quadrupole moment $\tilde{Q}$ (right inset)
  as functions of $M$.
  Bottom-right panel: Jordan-frame tidal ($\tilde{\lambda}$) and
  rotational ($\tilde{\lambda}^{\rm rot}$) Love numbers as functions
  of $M$.
\label{fig:structure}}
\end{center}
\end{figure*}
\begin{figure*}[t]
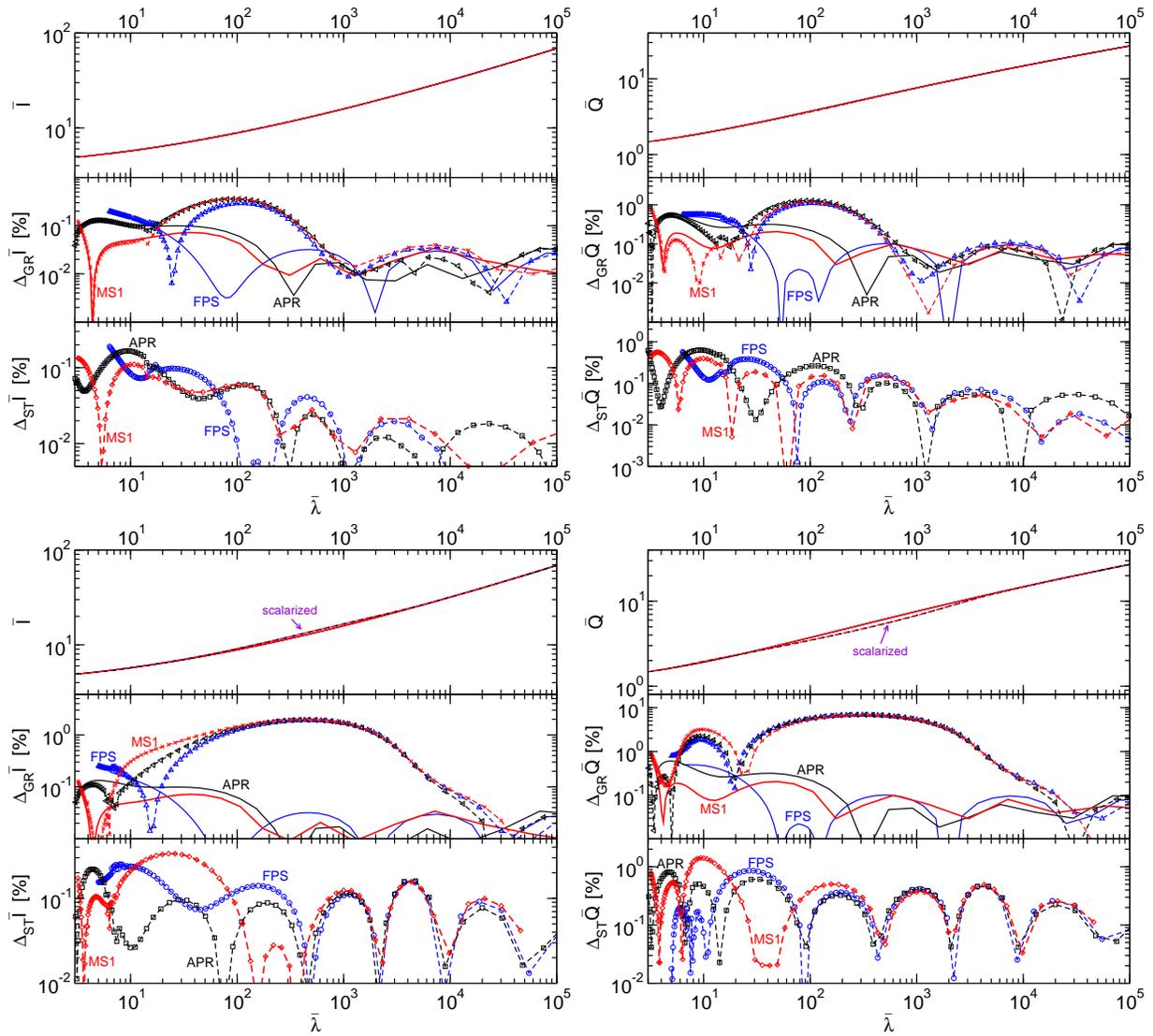

\begin{center}
\begin{tabular}{cc}
\epsfig{file=ILove_Jordan.eps,width=8cm,angle=0,clip=true}&
\epsfig{file=QLove_Jordan.eps,width=8cm,angle=0,clip=true}\\
\epsfig{file=ILove_large_Jordan.eps,width=8cm,angle=0,clip=true}&
\epsfig{file=QLove_large_Jordan.eps,width=8cm,angle=0,clip=true}
\end{tabular}
\caption{EOS-independent relations $\bar{I}(\bar{\lambda})$ (left) and
  $\bar{Q}(\bar{\lambda})$ (right). Solid linestyles refer to data in
  GR, dashed linestyles to data for scalarized stars. In each panel,
  the top inset shows the relation itself; the middle and bottom
  insets show deviations from universality, as measured by the
  residual $\Delta X= 100[X/X_{\rm fit}-1]$. $\Delta_{GR}X$ means that
  the universal relation is obtained by fitting only pure GR
  solutions; $\Delta_{ST}X$ means that the fit is obtained only from
  scalarized solutions. The top panels show that both residuals are
  always smaller than $2\%$, and typically smaller than $1\%$, for
  scalar-tensor theories that are marginally ruled out by binary
  pulsar observations. The bottom panels show that the residuals in
  $\bar{Q}(\bar{\lambda})$ can get as large as $\sim 10\%$ for
  theories that are already ruled out by experiment at more than
  $1\sigma$ confidence level.
\label{fig:ILoveQ}}
\end{center}
\end{figure*}
%
\section{Results}\label{sec:results}
For concreteness, in our numerical integrations we focus on a theory
defined by the action~\eqref{actionEinstein} with $V(\Phi)=0$ and
\begin{equation}
 A(\Phi)=e^{\frac{\beta}{2}\Phi^2}\,. \label{A}
\end{equation}

Isolated NSs in this theory were studied in
Refs.~\cite{Damour:1993hw,Damour:1996ke}, where it was shown that GR
solutions become energetically disfavored for sufficiently negative
values of $\beta$ due to a phase transition (``spontaneous
scalarization'') analogous to spontaneous magnetization in
ferromagnetism. Therefore in some regions of the parameter space the
theory admits stable NS configurations with nonvanishing scalar charge
(${q}\neq 0$).

In addition to the coupling parameter $\beta$, the theory is also
defined by the asymptotic value of the scalar field at infinity,
$\Phi_0^\infty$. Binary-pulsar observations~\cite{Freire:2012mg}
require $\beta/(4\pi)\gtrsim -4.5$, whereas the measurement of the
Shapiro time delay by the Cassini spacecraft~\cite{Bertotti:2003rm}
implies $\omega_{\rm BD}>4\times10^4$, where $\omega_{\rm BD}$ is
related to the asymptotic value of the scalar field through~\cite{Palenzuela:2013hsa}
\begin{equation}
 \Phi_0^\infty=\frac{2\sqrt{\pi}}{|\beta|\sqrt{3+2\omega_{\rm BD}}}\,.\label{defomega}
\end{equation}
Binary-pulsar constraints set even stronger bounds on $\Phi_0^\infty$
when $\beta\lesssim-2$ and, in fact, the upper bound on
$\Phi_0^\infty$ decreases very steeply as $\beta/(4\pi)\to
-4.5$~\cite{Freire:2012mg}.

Using nuclear-physics based tabulated EOSs, we have computed slowly
rotating NS configurations in this theory and extracted all relevant
quantities to second order in the NS angular momentum. As a further
test of our procedure, we have reproduced the results of
Refs.~\cite{Damour:1993hw,Damour:1996ke} for scalarized NSs to first
order in the spin. The second-order results presented below are
new. In our analysis we used three different EOSs covering a wide
range of stiffness, namely \texttt{FPS}, \texttt{APR}, and
\texttt{MS1} (cf. e.g. Ref.~\cite{Read:2008iy} for a discussion of the
models).

A summary of our findings is presented in Fig.~\ref{fig:structure}.
We perform numerical integrations in the Einstein frame, but all
physical quantities shown in Fig.~\ref{fig:structure} refer to the
physical (Jordan) frame, except for the mass which refers to the
Einstein-frame Arnowitt-Deser-Misner mass (see Appendix~\ref{app:EtoJ}
for details and for the relation between the two frames).  The figure
contains four panels. Each panel presents results for three models:
(i) GR solutions (solid lines); (ii) scalarized solutions where the
theory parameters are marginally excluded by binary pulsar
experiments\footnote{Recent unpublished observations of PSR J0348+0432
  seem to exclude the region $\beta/(4\pi)\lesssim -4.2$
  \cite{WexPC}. In order to maximize deviations from GR, we use very
  conservative parameters ($\beta/(4\pi)=-4.5$ and
  $\Phi_0^\infty=10^{-3}$) for a marginally excluded scalar-tensor
  theory. The new observational bounds, if confirmed, would only
  strengthen our conclusions.}, i.e. $\beta/(4\pi)=-4.5$ and
$\Phi_0^\infty=10^{-3}$ (dashed lines); (iii) scalarized solutions
where the theory parameters violate current experimental bounds at
more than $1\sigma$ confidence level~\cite{Freire:2012mg},
i.e. $\beta/(4\pi)=-6$ and $\Phi_0^\infty=10^{-3}$ (dash-dotted
lines).

Above some critical value of the central pressure $P_{c}$, the exact
value depending on the EOS, scalarized solutions coexist with their GR
counterpart. A linear perturbation analysis and numerical simulations
of stellar collapse show that the domain of existence of the
scalarized solutions coincides with the region where spherically
symmetric GR solutions are linearly unstable and spontaneously develop
a scalar charge~\cite{Harada:1997mr,Novak:1997hw,Pani:2010vc}. The
effects of scalarization are clear in the left inset of the top-left
panel of Fig.~\ref{fig:structure}, where we show the mass-radius
diagram for the GR branch and for two scalarized
theories. Rotationally induced mass
corrections (shown in the right inset) are sensibly theory-dependent.
The top-right panel shows the scalar charge (left inset) and
rotationally induced corrections to the scalar charge (right inset) as
functions of the stellar mass for scalarized solutions constructed
using different EOS models. Corrections to the scalar charge can be
very large, with $\delta \tilde{q}/\tilde{q}\sim 2$ for some values of
the mass. This is consistent with the findings of Doneva {\it et al.}
\cite{Doneva:2013qva}, who showed that rotation strengthens the
effects of scalarization: roughly speaking, the total energy of the
star must be large enough in order to scalarize, and scalarization is
favored in spinning stars because of the rotational contribution to
the total energy. The bottom-left panel shows that scalarization
affects the moment of inertia (left inset) and the quadrupole moment
(right inset). Finally, the bottom-right panel shows that tidal and
rotational Love numbers are nontrivially modified by scalarization,
with very large deviations in the case of theories that are already
ruled out by binary pulsar experiments.

Although all quantities to second order in the spin display large
modifications for different EOSs and also relative to GR, the behavior
of the dimensionless quantities~\eqref{ILoveQbar} turns out to be
much more universal.  In Fig.~\ref{fig:ILoveQ} we show the
$\bar{I}(\bar{\lambda})$ (left panels) and $\bar{Q}(\bar{\lambda})$
(right panels) relations for scalarized solutions. In the top panels,
scalarized solutions refer to a theory with $\beta/(4\pi)=-4.5$ and
$\Phi_0^\infty=10^{-3}$; in the bottom panels the theory parameters
are $\beta/(4\pi)=-6$ and $\Phi_0^\infty=10^{-3}$.

Let us first focus on the most relevant case, that of solutions that
are only marginally disfavored by experiment (top panels). The top
insets show six curves, corresponding to scalarized and nonscalarized
solutions for three different EOSs, but these curves are
indistinguishable on the scale of the plot: both in GR and in
scalar-tensor theories, the I-Love-Q relations display very small
deviations from universality. In general, the universal I-Love-Q
relation will depend on our assumption on the correct theory of
gravity: we can construct I-Love-Q relations either by fitting only
pure GR solutions (middle inset in each panel), or by fitting only
scalarized solutions (bottom inset). In the middle inset we show
deviations from ``pure-GR universality'' for stars in GR (continuous
lines) and for scalarized stars (dashed lines with
symbols). Deviations from universal relations are typically of the
order of $2\%$ or less for both $\bar{I}(\bar{\lambda})$ and
$\bar{Q}(\bar{\lambda})$.
Furthermore, the universal relations in experimentally viable
scalar-tensor theories are very close to their GR counterparts.

One could have expected a priori that universal relations in
scalar-tensor gravity would differ from those in GR, with larger
deviations for larger absolute values of the coupling parameter
$|\beta|$. The top panels of Fig.~\ref{fig:ILoveQ} show that, even for
a theory that is already marginally ruled out by binary-pulsar measurements,
the I-Love-Q relations agree with those in GR
within a few percent and, in fact, the deviation is comparable with
the spread between different EOS models \emph{within} GR.  In order to
assess the dependence on the coupling parameters, in the bottom panels
of Fig.~\ref{fig:ILoveQ} we show results for a theory with
$\beta/(4\pi)=-6$ and $\Phi_0^\infty=10^{-3}$, that is already
excluded by binary pulsar experiments at more than $1\sigma$
confidence level~\cite{Freire:2012mg}.  In this unrealistic case the
residual from the GR universal relation can be as large as $\sim10\%$
(cf. middle insets in the bottom panels Fig.~\ref{fig:ILoveQ}),
whereas \emph{within} the scalarized theory the
I-Love-Q relations are still nearly universal,
as shown by the small residuals in the lower insets of the bottom
panels.

For both scalar-tensor theories, the bottom insets highlight a rather
interesting point: if we consider scalar-tensor theory as the correct
theory of gravity, the deviations from a universal relation obtained
by fitting numerical data {\em within the theory} are always very
small. This means that the I-Love-Q relations are nearly universal,
independently of whether GR or scalar-tensor theory is the correct
theory of gravity. In other words, the universality is intimately tied
to universal properties {\em of matter},
and it is quite insensitive to the dynamics of strong-field gravity.

\begin{figure}[t]
\begin{center}
\begin{tabular}{c}
\epsfig{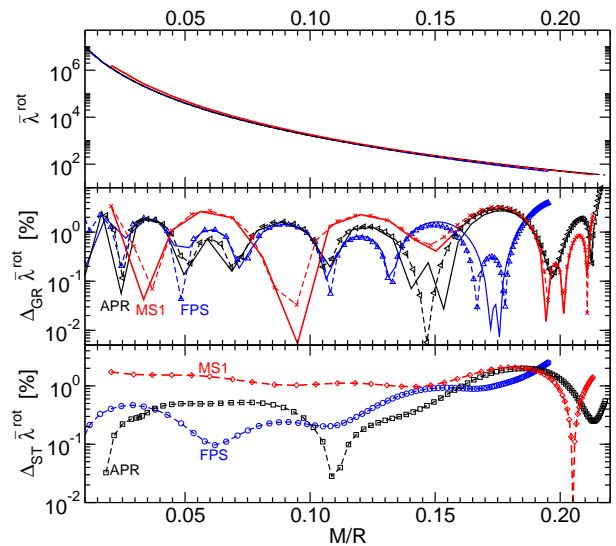}
\end{tabular}
\caption{Dimensionless rotational Love number $\bar{\lambda}^{\rm rot}
  \equiv\bar{I}^2\bar{Q}$ as a function of the compactness
  $M/R$ in a scalar-tensor theory defined by $A(\Phi)\equiv
  e^{\frac{\beta}{2}\Phi^2}$ and $V(\Phi)\equiv0$, for
  $\beta/(4\pi)=-4.5$, $\Phi_0^\infty=10^{-3}$ and for different
  tabulated EOS models. The residuals shown in the insets are defined
  as in Fig.~\ref{fig:structure}, and they are always smaller than a
  few percent. All quantities refer to the Jordan frame, $M$ is in solar-mass units whereas $R$ is in units of kilometers.
\label{fig:rotLove}}
\end{center}
\end{figure}

Finally, the dimensionless rotational Love number~\eqref{lambdarot} is
shown in Fig.~\ref{fig:rotLove} as a function of the stellar
compactness $M/R$ for a scalar-tensor theory that is only marginally
ruled out by binary pulsar experiments. As shown in the middle and
lower insets, the residuals from a universal fit are smaller than
$\sim 5\%$, even for scalarized solutions. An interesting fact is
that, contrarily to the cases shown in Figs.~\ref{fig:structure}
and~\ref{fig:ILoveQ}, the residuals of the scalarized solutions do not
increase relative to the GR solutions, i.e.  scalarization seems to
affect the dimensionless rotational Love number even less than other
quantities. A measurement of the NS mass and radius can be used to
infer the rotational Love number even if the underlying theory of
gravity is scalar-tensor theory.

In conclusion, the degeneracy between the I-Love-Q relations in GR and
in scalar-tensor theories that allow for scalarization is a nontrivial
fact. The degeneracy holds because of the tight experimental bounds
imposed on scalarization by current binary pulsar experiments, and it
is conceptually very different from the degeneracy observed in EiBI
theory~\cite{Sham:2013cya}. In that case, the degeneracy occurs
because the theory does not contain any extra degree of freedom with
respect to GR. As a consequence, perfect-fluid NS solutions in EiBI
can be mapped to GR solutions with a different
EOS~\cite{Delsate:2012ky}. The case discussed here is more
interesting, because scalar-tensor theories propagate an extra scalar
degree of freedom, so they are dynamically different from GR. Even at
the mathematical level, all the equations that define the I-Love-Q
relations depend explicitly on the background scalar field and, in
turn, on the scalar charge $q$. This result limits the prospects
of performing strong-field tests of GR using I-Love-Q relations. On
the plus side, it also means that astrophysical measurements of {\em
  any} of the three quantities ($\bar{I}$, $\bar{\lambda}$ or
$\bar{Q}$) can be used to infer the other two, quite independently of
assumptions on the EOS, as long as the underlying theory of gravity is
well constrained by weak-field or binary-pulsar experiments.

\section{Conclusions}\label{sec:conclusions}
We have presented a framework to construct slowly-rotating NSs in a
generic scalar-tensor theory of gravity, extracting all relevant
quantities to second order in the NS spin: mass, spin, scalar charge,
moment of inertia, spin-induced quadrupole, tidal and rotational Love
numbers.

We have focused on the simplest theory allowing for spontaneous
scalarization~\cite{Damour:1993hw,Damour:1996ke}, but our equations
(available online~\cite{webpages}) can be directly integrated in
\emph{any} scalar-tensor theory. In particular, our framework can be
used to study NSs in $f(R)$ gravity theories by virtue of
their equivalence with scalar-tensor
theories~\cite{Sotiriou:2008rp,DeFelice:2010aj}.

We have found that the nearly universal I-Love-Q relations that were
recently discovered in GR~\cite{Yagi:2013bca,Yagi:2013awa} are very
accurate (better than a few percent) for scalar-tensor theories that
allow for spontaneous scalarization within current experimental
bounds. Even for a theory that is already ruled out by observations,
the universal relations agree with their GR counterparts within $10\%$
or less, whereas for a theory that is only marginally viable the
deviations are lower than $2\%$, i.e. comparable to the dispersion due
to a different EOS within GR.

Our results imply that the simplest, best motivated and most-studied
extension of GR cannot be distinguished from Einstein's gravity using
tests based on the I-Love-Q triad (cf. Ref.~\cite{Sham:2013cya} for
another example).
On the other hand, our analysis tests the robustness of the I-Love-Q
relations against beyond-GR corrections, showing that the relations
derived in GR survive in scalar-tensor theories that are
phenomenologically viable. This suggests that a measurement of one
element of the I-Love-Q triad can be used to infer the remaining two
quantities within less than a few percent, even adopting a relatively
agnostic view on the behavior of gravity in the strong-curvature
regime, which remains experimentally unexplored to date.

Finally, for a given scalar-tensor theory the dispersion from
universality due to different EOSs is always smaller than a few
percent, quite independently of the coupling parameters appearing in
the action. This observation illustrates that the I-Love-Q relations
remain EOS-independent in scalar-tensor gravity, and it seems to
suggest that NS universal relations hinge more deeply on the
``extrinsic'', global properties of ultra-stiff matter, rather than on
the ``dynamical'' properties of the underlying gravitational theory.

In order to test these implications, it would be interesting to extend
our study to other scalar-tensor theories. Examples include: (i)
theories where the scalar field is massive, $V(\Phi)\sim m_s^2\Phi^2$,
which can evade weak-field tests~\cite{Alsing:2011er,Berti:2012bp} and
give rise to interesting strong-field effects, such as the existence
of floating orbits~\cite{Cardoso:2011xi}; (ii) tensor multi-scalar
theories \cite{Damour:1992we}; (iii) Horndeski
theory~\cite{Horndeski:1974wa,Sotiriou:2014jla}.

Another interesting avenue for future investigation is the extension
of our study to universal relations between high-order multipoles in
scalar-tensor theory.  Studies in GR~\cite{Yagi:2014bxa} show that
high-order multipole relations have larger spread than the original
I-Love-Q relations. Furthermore, high-order multipoles are harder to
measure than low-order multipoles. For these reasons it seems unlikely
that high-multipole relations will help in discriminating between
scalar-tensor theories with spontaneous scalarization and GR better
than the I-Love-Q relations. In any event, this is an interesting
possibility that should be explored.

It will also be interesting to extend the second-order in rotation
formalism developed here to other theories, such as Einstein-dilaton
Gauss-Bonnet gravity: first-order calculations were carried out
in~\cite{Pani:2009wy,Pani:2011gy,Pani:2011xm}, and recently extended
to slowly-rotating black holes at second order in
rotation~\cite{Ayzenberg:2014aka}, but (to the best of our knowledge)
second-order calculations of stellar structure were not reported in
the literature.

Finally, our results can be complemented and extended by constructing
fast-rotating NS solutions in scalar-tensor theories
(see~\cite{Doneva:2013qva} for work in this direction). This would
allow us to verify whether the I-Love-Q universality in scalar-tensor
theories is accurate enough for large rotation, as it seems to be in
GR~\cite{Doneva:2013rha,Pappas:2013naa,Chakrabarti:2013tca,Yagi:2014bxa}. In
this context, the results of our slow-rotation study can be used as a
benchmark and code test for full numerical solutions.

\begin{acknowledgments}
We are grateful to Michael Horbatsch, 
Hector Okada da Silva, George Pappas, Kent Yagi and Nico Yunes for constructive criticism
on an early draft of this paper.  PP acknowledges the kind hospitality
of the Harvard-Smithsonian Center for Astrophysics, where part of this
work has been initiated. This work was supported by the European
Community through the Intra-European Marie Curie contract
aStronGR-2011-298297 and IRSES grant NRHEP 295189
FP7-PEOPLE-2011-IRSES, and by FCT-Portugal through projects
IF/00293/2013 and CERN/FP/123593/2011.  EB is supported by NSF CAREER
Grant No.~PHY-1055103.
\end{acknowledgments}

\appendix
\begin{widetext}
\section{Hartle-Thorne second-order equations for generic scalar-tensor theories}\label{app:eqs}
In this Appendix we present the field equations of a slowly-rotating,
perfect-fluid star to second order in the angular momentum in
scalar-tensor theories of gravity. We set $V(\Phi)\equiv0$ for
simplicity, but a more general form of the equations with $V(\Phi)\neq
0$ can be found in a publicly available {\scshape Mathematica}
notebook~\cite{webpages}, together with the procedure to integrate the
equations numerically (as explained in the main text).

Given the decomposition~\eqref{metric}--\eqref{rho} in the Einstein
frame, the zeroth-order quantities are described by the following
modified Tolman-Oppenheimer-Volkoff equations:

\begin{eqnarray}
 m'&=& 4 \pi  r \left(r A^4 \rho_0 +\frac{1}{2} (r-2 m) \Phi_0'^2\right)\,,\\
 \nu '&=& \frac{2 m+8 \pi  r^3 A^4 P_0}{r(r-2m)}+4 \pi  r \Phi_0'^2\,,\\
 P_0'&=& -(P_0+\rho_0 ) \left(\frac{m+4 \pi  r^3 A^4 P_0}{r(r-2m)}+\frac{A' \Phi_0'}{A}+2 \pi  r \Phi_0'^2\right)\,,\\
 \Phi_0''&=& \frac{r^2 A^3 (\rho_0-3 P_0 ) A'+2 \left[m-r+2 \pi  r^3 A^4 (\rho_0-P_0 )\right] \Phi_0'}{r (r-2 m)}\,.\nn\\
\end{eqnarray}
The only first-order quantity is $\bar\omega=\Omega-\omega$, which is described
by a second-order ODE:
\begin{eqnarray}
 \bar{\omega}''&=& \frac{1}{r (r-2 m)}4 \left[(r-2 m) \left(\pi  r^2  \Phi_0'^2-1\right) \bar{\omega}'
 +\pi  r^2 A^4 (P_0+\rho_0 ) \left(4 \bar{\omega}+r \bar{\omega}'\right)\right]\,. \label{eqomega}
\end{eqnarray}
To ${\cal O}(\epsilon^2)$, we obtain five first-order ODEs for $p_0$, $m_0$, $h_0$, $v_2$ and $h_2$,
\begin{eqnarray}
 p_0'&=& \frac{e^{-\nu }}{3 A^2 (r-2 m)} \left[2 r A^2 \bar{\omega} \left(\bar{\omega} \left(r-3 m-4 \pi  r^3 A^4 P_0-2 \pi  r^2 (r-2 m) \Phi_0'^2\right)+r (r-2 m) \bar{\omega}'\right)\right.\nn\\
 &&\left.-3 e^{\nu } (r-2 m) \left(A^2 h_0'-\phi_0 A'^2 \Phi_0'+A \left(A' \phi_0'+\phi_0 \Phi_0' A''\right)\right)\right]\,,\\
 m_0'&=& \left[e^{-\nu } r \left(128 \pi ^2 r^6 A^9 P_0 (P_0+\rho_0 ) \bar{\omega}^2+768 \pi ^2 e^{\nu }  r^4 A^8 P_0 \rho_0  \phi_0 A'+192\pi  e^{\nu }  r^2 A^3 (r-2 m) \rho_0  \phi_0 A'^2 \Phi_0'\right.\right.\nn\\
 &&\left.\left.+32 \pi  r A^4 A' \left(r^3 (r-2 m) (P_0+\rho_0 ) \bar{\omega}^2 \Phi_0'+6 e^{\nu } \rho_0  \phi_0 \left(m+2 \pi  r^2 (r-2 m) \Phi_0'^2\right)\right)\right.\right.\nn\\
 &&\left.\left.+r (r-2 m) A' \Phi_0' \left(-48 \pi e^{\nu }  \Phi_0' \left(-(r-2 m) \phi_0'+m_0 \Phi_0'\right)+r^2 (r-2 m) \bar{\omega}'^2\right)\right.\right.\nn\\
 &&\left.\left.+A \left(m+2 \pi  r^2 (r-2 m) \Phi_0'^2\right) \left(-48\pi e^{\nu }   \Phi_0' \left(-(r-2 m) \phi_0'+m_0 \Phi_0'\right)+r^2 (r-2 m) \bar{\omega}'^2\right)\right.\right.\nn\\
 &&\left.\left.+4 \pi  r^2 A^5 \left(8 r (P_0+\rho_0 ) \bar{\omega}^2 \left(m+2 \pi  r^2 (r-2 m) \Phi_0'^2\right)+12 e^{\nu } \left(-(r-2 m) p_0 \rho_0 '+4 \pi  r P_0 \Phi_0' \left((r-2 m) \phi_0'-m_0 \Phi_0'\right)\right)  \right.\right.\right.\nn\\
 &&\left.\left.\left.  +r^3 (r-2 m) P_0 \bar{\omega}'^2\right)\right)\right]\nn\\
 &&\times\left[12 \left(4 \pi  r^3 A^5 P_0+r (r-2 m) A' \Phi_0'+A \left(m+2 \pi  r^2 (r-2 m) \Phi_0'^2\right)\right)\right]^{-1}\,,
\end{eqnarray}
\begin{eqnarray}
 h_0'&=& -\frac{e^{-\nu } }{12 (r-2 m)^2}\left(-12 e^{\nu } \left(m_0 \left(1+8 \pi  r^2 A^4 P_0\right)+4 \pi  r (r-2 m) \left(r A^3 \left(A p_0 (P_0+\rho_0 )+4 P_0 \phi_0 A'\right)+(r-2 m) \phi_0' \Phi_0'\right)\right) \right.\nn\\
 &&\left. +r^3 (r-2 m)^2 \bar{\omega}'^2\right)\,,\label{eq:h0}\\
 v_2'&=& \frac{1}{6 r (r-2 m)}\left(12 h_2 \left(-m-4 \pi  r^3 A^4 P_0-2 \pi  r^2 (r-2 m) \Phi_0'^2\right)+e^{-\nu } \left(-48\pi e^{\nu }   r (r-2 m) \phi_2 \Phi_0'\right.\right.\nn\\
 &&\left.\left. +r^3 \left(r-m+4 \pi  r^3 A^4 P_0+2 \pi  r^2 (r-2 m) \Phi_0'^2\right) \left(16 \pi  r A^4 (P_0+\rho_0 ) \bar{\omega}^2+(r-2 m) \bar{\omega}'^2\right)\right)\right)\,,
 \end{eqnarray}
\begin{eqnarray}
 h_2'&=& \left[e^{-\nu } \left(-24 e^{\nu } h_2 \left(r m-m^2-2 \pi  r^4 A^4 P_0+12 \pi  r^3 A^4 m P_0+16 \pi ^2 r^6 A^8 P_0^2-2 \pi  r^4 A^4 \rho_0 +4 \pi  r^3 A^4 m \rho_0 \right.\right.\right.\nn\\
 &&\left.\left.\left. +2 \pi  r^3 (r-2 m) \left(1+8 \pi  r^2 A^4 P_0\right) \Phi_0'^2+4 \pi ^2 r^4 (r-2 m)^2 \Phi_0'^4\right)+r \left(-24 e^{\nu } (r-2 m) v_2+512 \pi ^3 r^9 A^{12} P_0^2 (P_0+\rho_0 ) \bar{\omega}^2\right.\right.\right.\nn\\
 &&\left.\left.\left. -48 \pi e^{\nu }   r^2 A^3 (r-2 m) (3 P_0-\rho_0 ) \phi_2 A'+32 \pi ^2 r^6 A^8 P_0 \left(8 (P_0+\rho_0 ) \bar{\omega}^2 \left(m+2 \pi  r^2 (r-2 m) \Phi_0'^2\right)+r^2 (r-2 m) P_0 \bar{\omega}'^2\right)\right.\right.\right.\nn\\
 &&\left.\left.\left. +(r-2 m) \left(-48\pi  e^{\nu }  \Phi_0' \left(2 \phi_2 \left(r-m+2 \pi  r^2 (r-2 m) \Phi_0'^2\right)+r (r-2 m) \phi_2'\right)\right.\right.\right.\right.\nn\\
 &&\left.\left.\left.\left. +r^2 \left(-r^2+2 m (r+m)+8 \pi  r^2 (r-2 m) (r-m) \Phi_0'^2+8 \pi ^2 r^4 (r-2 m)^2 \Phi_0'^4\right) \bar{\omega}'^2\right)\right.\right.\right.\nn\\
 &&\left.\left.\left. +16 \pi  r^3 A^4 \left(\rho_0  \bar{\omega}^2 \left(r^2+2 m (m-r)+8 \pi  r^2 (r-2 m) (r-m) \Phi_0'^2+8 \pi ^2 r^4 (r-2 m)^2 \Phi_0'^4\right)\right.\right.\right.\right.\nn\\
 &&\left.\left.\left.\left. +P_0 \left(\bar{\omega}^2 \left(r^2+2 m (m-r)+8 \pi  r^2 (r-2 m) (r-m) \Phi_0'^2+8 \pi ^2 r^4 (r-2 m)^2 \Phi_0'^4\right)\right.\right.\right.\right.\right.\nn\\
 &&\left.\left.\left.\left.\left.+(r-2 m) \left(-24 \pi e^{\nu }  \phi_2 \Phi_0'+r^2 \left(m+2 \pi  r^2 (r-2 m) \Phi_0'^2\right) \bar{\omega}'^2\right)\right)\right)\right)\right)\right]\nn\\
 &&\times\left[12 r (r-2 m) \left(m+4 \pi  r^3 A^4 P_0+2 \pi  r^2 (r-2 m) \Phi_0'^2\right)\right]^{-1}\,, 
\end{eqnarray}
%
%
and two second-order ODEs for $\phi_0$ and $\phi_2$, that we write
schematically as
\begin{eqnarray}
 \phi_0''+C_1 \phi_0'+C_0\phi_0&=&S_1 \,,\\
 \phi_2''+C_1 \phi_2'+D_0\phi_2&=&S_2 \,,
\end{eqnarray}
where the radial coefficients $C_1$, $C_0$ and $D_0$, as well as the
source terms $S_i$, are lengthy and unenlightening; their form can be
found in the {\scshape Mathematica} notebook~\cite{webpages}.

Using the other field equations, the right-hand side of
Eq.~\eqref{eq:h0} can be written as a total derivative and integrated,
with the result
\begin{equation}
 h_0(r)={\rm constant}-p_0+\frac{r^2  \bar{\omega}^2}{3} e^{-\nu } -\phi_0 \frac{A'}{A}\,.
\end{equation}
This expression reduces to Eq.~(17b) in Ref.~\cite{Hartle:1968si} in
the GR limit $A(\Phi)\equiv1$.
Finally, the functions $p_2$ and $m_2$ are algebraically related to
the others through
\begin{eqnarray}
 p_2&=& -h_2-\frac{r^2 \bar{\omega}^2}{3} e^{-\nu }-\phi_2\frac{A'}{A}\,,\\
 m_2&=& \frac{e^{-\nu }}{6}  (r-2 m) \left[r^3 \left(16 \pi  r A^4 (P_0+\rho_0 ) \bar{\omega}^2
 +(r-2 m) \bar{\omega}'^2\right)-6 e^{\nu } h_2\right]\,.
\end{eqnarray}

The equations for $v_2$, $h_2$, $\phi_0$ and $\phi_2$ are
nonhomogeneous. As explained in the main text, the appropriate
boundary conditions can be imposed with the help of the homogeneous
equations, along the lines of the GR case~\cite{Hartle:1968si}. We
denote the solutions of the homogeneous equations as $v_2^{(h)}$,
$h_2^{(h)}$, $\phi_0^{(h)}$ and $\phi_2^{(h)}$, respectively. In
reduced first-order form, we need to solve a system of 16 coupled ODEs
plus six homogeneous first-order ODEs for $v_2^{(h)}$, $h_2^{(h)}$,
$\phi_0^{(h)}$ and $\phi_2^{(h)}$ . The system can be written
schematically as in Eq.~(\ref{system}).

Note that the equations at first order in rotation are a particular
case of those presented in Ref.~\cite{Pani:2011xm}\footnote{We note
  here that the field equation in Ref.~\cite{Pani:2011xm} contain some
  typos when $A\neq0$, which are corrected in this paper and in the
  {\scshape Mathematica} notebook~\cite{webpages}.}. Here we have
extended the analysis to second order, focusing on generic
scalar-tensor theories.

\section{Transformations of the physical quantities to the Jordan frame}
\label{app:EtoJ}
The field equations that we integrate numerically are derived from the
Einstein-frame action of Eq.~\eqref{actionEinstein}. In the Einstein
frame, matter fields are nonminimally coupled with the conformal
Einstein metric $g_{\mu\nu}$. However, laboratory clocks and rods
measure the ``physical'' metric $\tilde{g}_{\mu\nu}$
that appears in the Jordan-frame action of Eq.~\eqref{actionST}. In
this Appendix we explicitly give the transformations relating the
macroscopic properties characterizing NSs in the two frames.

Since the moment of inertia, the quadrupole moment and the tidal Love
number all depend on the fall-off of the metric at large distances,
once the asymptotic behavior of the Einstein-frame metric $g_{\mu\nu}$
and of the scalar field $\Phi$ are known, the Jordan-frame quantities
can be easily computed from the asymptotic behavior of the
Jordan-frame metric
\begin{equation}
\tilde{g}_{\mu\nu}=A(\Phi)^2 g_{\mu\nu}\,, \label{conformal}
\end{equation}
where $\Phi$ is defined in Eq.~\eqref{scalardef}, and we recall that
tilded quantities refer to the Jordan frame.

Because we imposed asymptotic flatness on the Einstein-frame metric,
the conformal transformation \eqref{conformal} yields
$\tilde{g}_{\mu\nu}\to A(\Phi_0^\infty)^2 \eta_{\mu\nu}$ at
infinity. In order for the Jordan-frame metric to be also
asymptotically Minkowskian, we can simply rescale the time and radial
coordinates: $\tilde{t}=A(\Phi_0^\infty) t$ and
$\tilde{r}=A(\Phi_0^\infty) r$. Note that, for phenomenologically
viable values of $\Phi_0^\infty$, $A(\Phi_0^\infty)\simeq 1$ to a very
good approximation, so this rescaling is practically negligible.

In addition, in scalar-tensor theories the effective gravitational
constant $\tilde{G}$ (as measured by a Cavendish-like experiment) is
not necessarily the same as the ``bare'' constant $G$ appearing in
Eq.~\eqref{actionST} (recall that we set $G=1$ in our Einstein-frame
integrations). For the theory considered in the main text, where
$V(\Phi)=0$ and $A(\Phi)$ is given by Eq.~\eqref{A}, the relation
between these two quantities reads (see
e.g.~\cite{Palenzuela:2013hsa})
\begin{equation}
 \tilde{G}=e^{\beta \gamma^2}\left[G +\frac{\beta^2\gamma^2}{4\pi}\right]\sim  G e^{\beta \gamma^2}\,,\label{effectiveG}
\end{equation}
where for ease of notation we defined $\gamma=\Phi_0^\infty$, and in
the last step we have neglected the second term in square brackets,
because it is negligible in the phenomenologically viable region of
the $(\beta,\gamma)$ parameter space.  Thus, in the physical frame
some coefficients of the large-distance expansion of the metric
$\tilde{g}_{\mu\nu}$ must be rescaled. For example, the Jordan-frame
Arnowitt-Deser-Misner mass $\tilde{M}$ is obtained by comparing the
asymptotic expansion of Eq.~\eqref{conformal} with
$1/\tilde{g}_{rr}\to 1-2 \tilde{G} \tilde{M}/\tilde{r}$ at large
distances. By applying a similar procedure to the other components of
$\tilde{g}_{\mu\nu}$ we obtain
%
\begin{eqnarray}
 \tilde{M}&=& e^{-\frac{\beta}{2}\gamma^2} (M+\beta  \gamma {C}) \,,\\
  \tilde{J}&=&  J \,,\\
 \tilde{{q}}&=& -2\beta \gamma e^{-\frac{\beta}{2}\gamma^2} q  \,,\\
 \tilde{I}&=&  I \,,\\
 \tilde{Q}&=& e^{\frac{\beta}{2}\gamma^2} (Q+ \beta  \gamma {Q_s})  \,,
\end{eqnarray}
and
%
\begin{eqnarray}
 \tilde{\lambda}&=& e^{\frac{3\beta}{2}\gamma^2} \lambda +e^{\frac{3\beta}{2}\gamma^2}\frac{C}{135} \left[30  M \beta \lambda_s  \gamma+48 M^4 \beta  \gamma-4 M^3 C \left(66 \pi +5 \beta  \left(1+2 \beta  \gamma^2\right)\right)-6 M^2 \beta  \gamma C^2 \left(5 \beta  \left(3+2 \beta  \gamma^2\right)-24 \pi\right)\right.\nn\\
 &&\left.+2 M C^3 \left(26 \pi ^2+20 \pi  \beta  \left(1+2 \beta  \gamma^2\right)+5 \beta ^2 \left(3+4 \beta  \gamma^2 \left(3+\beta  \gamma^2\right)\right)\right) \right.\nn\\
 &&\left.+\beta  \gamma C^4 \left( +20 \pi  \beta  \left(3+2 \beta  \gamma^2\right)+3 \beta ^2 \left(15+4 \beta  \gamma^2 \left(5+\beta  \gamma^2\right)\right)-52 \pi^2\right) \right] \,.
\end{eqnarray}
Note that the moment of inertia is the same in both frames. This
follows from the fact that $\tilde{J}= J$ and that also the fluid
angular velocity, $\Omega$, is the same in both frames: $\Omega\equiv
u^\varphi/u^t=(A(\Phi)\tilde{u}^\varphi)/(A(\Phi)\tilde{u}^t)$. The
transformation of the tidal Love number is more complex than the
others because it depends on the expansion~\eqref{H0inf}, whose
subleading terms mix --~through Eq.~\eqref{conformal}~-- with the
fall-off of the scalar field.

Since the issue about which frame should be considered ``physical'' in
scalar-tensor theories is still debated (see
e.g.~\cite{Faraoni:1999hp,Sotiriou:2007zu}) we have computed all
quantities in both frames, with very similar results. All of the
numerical results presented in the main text refer to the (measurable)
Jordan-frame quantities, except for the mass. In theories that violate
the strong equivalence principle the notion of mass is
subtle. Following previous work (see
e.g.~\cite{Shibata:2013pra,Doneva:2013qva}), here we decided to
present the Arnowitt-Deser-Misner mass in the Einstein frame $M$,
which coincides with the so-called tensor mass and has several
desirable properties: it is positive definite, it decreases
monotonically under gravitational-wave emission and it is well defined
even for dynamical spacetimes~\cite{Lee:1974pt}.
\end{widetext}

\bibliography{biblio}  
\end{document}